\documentclass[lettersize,journal]{IEEEtran}
\usepackage{amsmath,amsfonts}
\usepackage{enumitem} 
\usepackage{algorithmic}
\usepackage{array}
\usepackage[caption=false,font=normalsize,labelfont=sf,textfont=sf]{subfig}
\usepackage{textcomp} 
\usepackage{stfloats}
\usepackage{url}
\usepackage{verbatim}
\usepackage{graphicx}
\def\BibTeX{{\rm B\kern-.05em{\sc i\kern-.025em b}\kern-.08em
    T\kern-.1667em\lower.7ex\hbox{E}\kern-.125emX}}
\usepackage{balance}
\begin{document}
\title{Mesquite MoCap: Democratizing Real-Time Motion Capture with Affordable, Bodyworn IoT Sensors and WebXR SLAM}

\author{Poojan Vanani, 
        Darsh Patel, 
        Danyal Khorami, 
        Siva Munaganuru, 
        Pavan Reddy, 
        Varun Reddy, 
        Bhargav Raghunath, 
        Ishrat Lallmamode,
        Romir Patel,
        Assegid Kidan\'e, 
        and Tejaswi Gowda


\thanks{Poojan Vanani is with [Arizona State University, Tempe, AZ, USA].}
\thanks{Darsh Patel is with [Arizona State University, Tempe, AZ, USA].}
\thanks{Danyal Khorami is with [Arizona State University, Tempe, AZ, USA].}
\thanks{Siva Munaganuru is with [Arizona State University, Tempe, AZ, USA].}
\thanks{Pavan Reddy is with [Arizona State University, Tempe, AZ, USA].}
\thanks{Varun Reddy is with [Arizona State University, Tempe, AZ, USA].}
\thanks{Bhargav Raghunath is with [Arizona State University, Tempe, AZ, USA].}
\thanks{Ishrat Lallmamode is with [Arizona State University, Tempe, AZ, USA].}
\thanks{Romir Patel is with [Basha High School, Chandler, AZ, USA].}
\thanks{Assegid Kidan\'e is with [Arizona State University, Tempe, AZ, USA].}
\thanks{Tejaswi Gowda is with [Arizona State University, Tempe, AZ, USA].}%
}

\markboth{IEEE Internet of Things Journal, Draft}%
{How to Use the IEEEtran \LaTeX \ Templates}

\maketitle

\begin{abstract}
Motion capture remains costly and complex to deploy, limiting use outside specialized laboratories. We present Mesquite, an open-source, low-cost inertial motion-capture system that combines a body-worn network of 15 IMU sensor nodes with a hip-worn Android smartphone for position tracking. A low-power wireless link streams quaternion orientations to a central USB dongle and a browser-based application for real-time visualization and recording. Built on modern web technologies—WebGL for rendering, WebXR for SLAM, WebSerial and WebSockets for device and network I/O, and Progressive Web Apps for packaging—the system runs cross-platform entirely in the browser. In benchmarks against a commercial optical system, Mesquite achieves mean joint-angle error of 2--5 degrees while operating at approximately 5\% of the cost. The system sustains 30~frames per second with end-to-end latency under 15ms and a packet delivery rate of at least 99.7\% in standard indoor environments. By leveraging IoT principles, edge processing, and a web-native stack, Mesquite lowers the barrier to motion capture for applications in entertainment, biomechanics, healthcare monitoring, human--computer interaction, and virtual reality. We release hardware designs, firmware, and software under an open-source license (GNU GPL).
\end{abstract}

\begin{IEEEkeywords}
  Motion Capture, Inertial Measurement Unit, Real-Time Tracking, Open-Source Hardware, Affordable Technology, Biomechanics, Animation, Virtual Reality, Internet of Things, Edge Computing, Sensor Networks, Human-Computer Interaction, Wearable Technology, SLAM, Web Technologies.
\end{IEEEkeywords}

\section{Introduction}

\IEEEPARstart{M}{otion capture} (MoCap) technology enables the recording, replay, and analysis of human movement, with applications spanning biomechanical research, clinical rehabilitation, animation, and interactive entertainment \cite{kitagawa2020}. Traditional optical motion capture systems provide high-precision data but require dedicated studio spaces, optimal lighting conditions, and investments often exceeding \$100,000 \cite{meng2021}. While IMU-based commercial alternatives offer portability, they remain costly due to proprietary hardware and software ecosystems \cite{riener2016}.

These constraints create significant accessibility barriers for many potential users, including small research laboratories, educational institutions, healthcare providers, and independent developers. Users seeking affordable alternatives often sacrifice customization and specificity, forcing them to compromise on both affordability and functionality. The limitations of existing motion capture systems stem from their reliance on expensive proprietary hardware and software, which restricts adaptability \cite{cheung2016}. Furthermore, the complexity of setup, calibration, and the need for dedicated spaces can deter users from fully utilizing the technology. As a result, many researchers and practitioners are left without effective tools for capturing and analyzing human movement in real-world environments.

The Internet of Things (IoT) paradigm \cite{atzori2010}, along with Body Worn Sensor Networks (BWSNs) \cite{pantelopoulos2010survey}, offers a promising approach to address these limitations through distributed sensing, edge computing, and wireless data aggregation. By leveraging advances in microelectromechanical systems (MEMS) \cite{foxlin2005}, low-power wireless microcontrollers, and emerging web standards (such as WebSerial \cite{webserial}, WebSockets \cite{websockets}, WebXR \cite{webxr}, WebGL \cite{webgl}, and Progressive Web Apps \cite{pwa}), it becomes possible to create affordable motion capture systems without compromising performance.

This paper presents Mesquite, an IoT-based motion capture system that utilizes a wireless network of inertial sensors and a consumer smartphone to track human movement. The system implements an independent network architecture, enabling real-time data transmission, browser-based visualization, and a record-replay pipeline. Our contributions include: a low-cost (under \$1000), open-source wireless sensor network design for human motion capture using commodity hardware; a robust network design optimized for real-time sensor data transmission with minimal latency; real-time kinematic reconstruction; comprehensive performance evaluation comparing accuracy with commercial systems; and demonstration of data integration across multiple domain toolchains.

The remainder of this paper is organized as follows: Section II reviews related work in IMU-based motion capture, IoT sensor networks, and wearable position and orientation estimation. Section III details the system architecture, covering hardware, network design, software components, data processing, and the record-and-reuse pipeline. Section IV presents performance evaluation and validation results. Section V discusses applications, followed by future directions in Section VI and conclusions in Section VII.

\section{Background and Related Work}
\subsection{IMU-Based Motion Capture}

Inertial Measurement Unit (IMU)-based motion capture has gained significant attention as a flexible and cost-effective alternative to traditional optical systems, which typically require complex setups and controlled environments. Unlike optical systems which rely on cameras and markers, IMU-based systems offer greater mobility, reduced infrastructure requirements, and the potential for use in unconstrained environments \cite{filippeschi2017}. Commercial solutions like Xsens MVN \cite{roetenberg2019}, Perception Neuron \cite{perception2021}, and Rokoko Smartsuit Pro \cite{rokoko} have demonstrated the viability of these approaches for full-body motion capture in entertainment, sports, and rehabilitation contexts. However, their high costs remain a barrier to broader adoption, particularly in educational, hobbyist, or low-resource research settings.

To address this, several research efforts have explored low-cost alternatives. Smartphone-based systems \cite{gowing2014, chen2018smartphone} and custom-built platforms using wireless microcontrollers \cite{tang2015, liu2020} aim to lower the cost of entry but often trade off extensibility, ease and mature toolchains that integrate with existing software. Mesquite addresses these issues by exporting both raw data and the industry standard Biovision Hierarchy(BVH) format\cite{meredith2001motion} which allows for easy integration with existing animation and motion capture software. Recent advancements in MEMS (Micro-Electro-Mechanical Systems) technology have improved the sensitivity, size, and cost of consumer-grade IMUs, making them suitable for precise motion tracking when combined with effective sensor fusion algorithms\cite{madgwick2011, foxlin2005}. These algorithms enable real-time orientation estimation by fusing data from accelerometers, gyroscopes, and magnetometers. These operations are increasingly performed on the sensor node itself, reducing the need for high-bandwidth wireless communication by using on-board motion processing\cite{roetenberg2019}.

Despite these improvements, existing low-cost solutions continue to face critical limitations. These include drift accumulation over time due to integration errors, restricted wireless communication architectures that hinder multi-sensor synchronization, and a general lack of robust, openly available software and hardware platforms that can be customized for application-specific needs \cite{filippeschi2017, pfeil2016}. Moreover, few systems offer comprehensive real-time visualization or support real time streaming out-of-the-box.

The core problem of IMU based mocap is calculating the position and orientation of the user in 3D space. While IMUs can provide accurate local motion data, they often struggle with drift and cumulative errors over time, making it difficult to maintain an accurate global reference frame. This is particularly challenging in dynamic environments where the user may move through complex trajectories or interact with physical objects. To solve this Mesquite can use any of the many implementations of Simultaneous Localization and Mapping (SLAM) techniques, which use computer vision to establish a global reference frame based on the user's environment. This approach allows for accurate tracking of the user's position and orientation in 3D space, even in dynamic environments. The system uses WebXR World Mapping \cite{webxr, webxr-wm} to establish an absolute spatial frame, which is then used to track the user's hip position. This approach allows for accurate tracking of the user's position in 3D space, even in dynamic environments, while the rest of the body is tracked using 15 body-worn IMUs for orientation, and joint positions are estimated through forward kinematics.

\subsection{IoT Sensor Networks for Human Motion Analysis}

The Internet of Things (IoT) paradigm has catalyzed innovative approaches to sensor deployment and distributed data collection for human motion analysis. In particular, Body Sensor Networks (BSNs)—wireless networks of wearable sensors—have been extensively developed for a range of applications including healthcare monitoring \cite{pantelopoulos2010survey}, physical rehabilitation \cite{yang2014}, activity recognition \cite{lara2013survey}, and sports performance analysis \cite{baca2010sensor}. These systems typically consist of multiple miniaturized wireless sensor nodes attached to various parts of the body, communicating with a central aggregator or hub that performs data fusion and higher-level analysis.

Several wireless communication protocols have been employed in BSNs, including Bluetooth Classic, Bluetooth Low Energy (BLE), ZigBee, Wi-Fi, and proprietary RF solutions \cite{otanes2019}. Each protocol involves trade-offs between power consumption, data rate, communication range, and device complexity. For example, BLE is widely used due to its low power profile, but it may not support the high data throughput from multiple nodes required for high-frequency motion capture involving multiple sensors \cite{milenkovic2006wireless}.

A major challenge in motion capture applications is the requirement for high sampling rates (typically 30–120 Hz), low-latency data transmission, and tight synchronization across multiple sensors. Many existing BSNs are optimized for long battery life rather than real-time performance, which limits their suitability for dynamic motion capture tasks such as full-body tracking in eXtended Reality (XR) and biomechanical analysis \cite{park2010wireless}.

To address these limitations, edge computing approaches have been proposed, in which preliminary data processing—such as sensor fusion, noise filtering, or compression—is performed on the sensor node or a nearby edge device before transmitting results to the central system. This strategy can reduce wireless bandwidth requirements, mitigate latency, and improve the system’s overall responsiveness \cite{rahmani2018edge}. Platforms such as Fog computing and lightweight embedded AI models are beginning to bridge the gap between real-time responsiveness and low-power operation in distributed sensing environments \cite{pham2021iot}. In our case, the Mesquite system leverages ICM20948 IMUs with integrated Digital Motion Processors (DMPs) to perform onboard sensor fusion, significantly reducing the amount of data transmitted over the network and enabling real-time performance in resource-constrained environments.

Despite these advances, integrating such techniques into affordable and scalable motion capture systems remains a significant challenge. Most existing solutions are either cost-prohibitive or too application-specific, lacking modularity and adaptability. The Mesquite system aims to address this gap by leveraging low-cost, Wi-Fi-enabled microcontrollers, synchronized real-time data acquisition, and real-time web-based rendering, providing a flexible BWSN IoT platform suitable for a wide range of motion analysis applications.

\subsection{Position and Orientation Estimation}
Accurate estimation of human body position and orientation is a fundamentally difficult problem, whether using optical systems, inertial sensors, or vision-based ML methods. Human motion involves complex dynamics, occlusion, and high degrees of freedom, all of which introduce challenges for real-time tracking. When using Inertial Measurement Units (IMUs), these difficulties are compounded by issues such as sensor drift, bias, environmental noise, and the lack of global spatial references.

IMU-based tracking systems rely on sensor fusion algorithms that combine data from accelerometers, gyroscopes, and magnetometers to estimate orientation. Algorithms such as the Kalman filter and its variants, including the Extended Kalman Filter (EKF), are widely used to optimally estimate the system state from noisy measurements \cite{welch2006}. Due to the limitations of Euler angles (e.g., gimbal lock), quaternion-based representations are typically employed for robust and continuous orientation tracking \cite{shoemake1985}.

Recent advances in sensor hardware, such as the BNO08x (Bosch)~\cite{boschbno} and ICM-20948 (TDK)~\cite{tdk20948}, have made it feasible to perform onboard sensor fusion using integrated motion processors. This not only improves accuracy but also reduces wireless data transmission overhead, a critical factor for real-time applications and networked sensor systems \cite{roetenberg2019, boschbno}.

\begin{figure*}[!t]
  \centering
  \includegraphics[width=\textwidth]{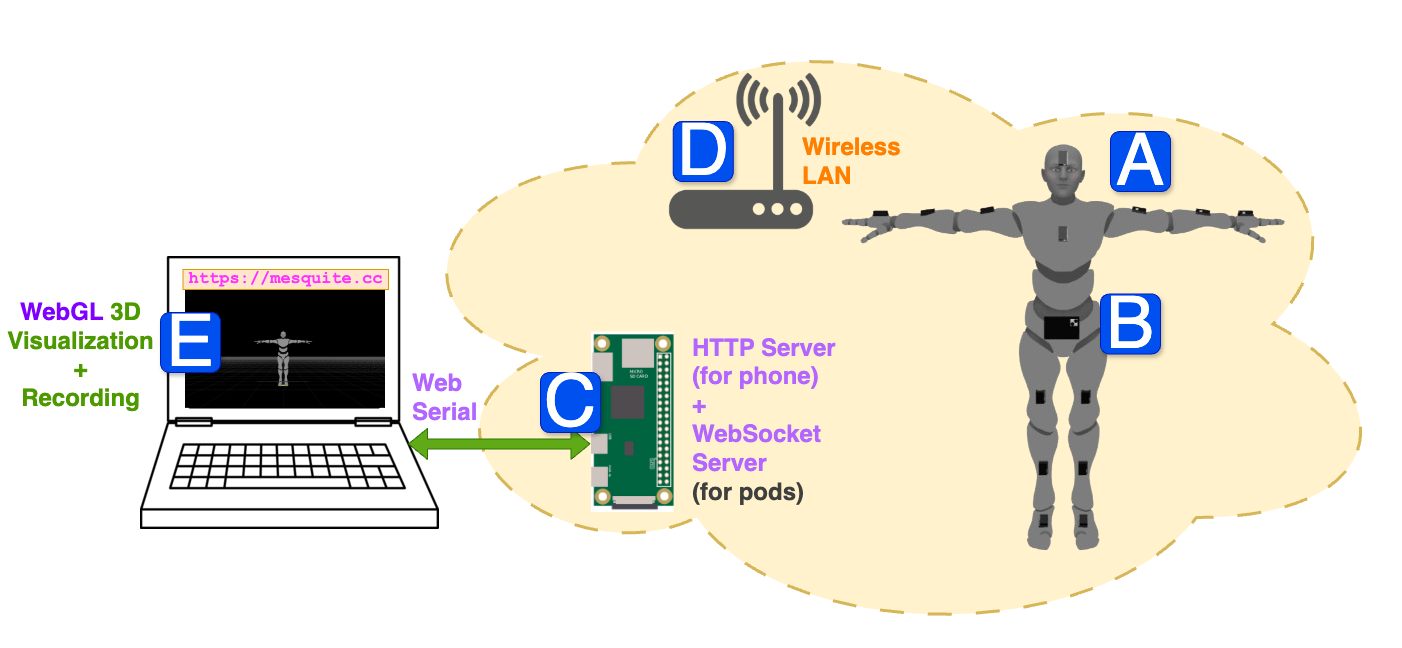}
  \caption{Overview of the Mesquite motion capture system showing (a)wearable sensor node (pod) placement on the human body. (b) Hipworn smartphone for spatial anchoring (c) Network architecture with a Raspberry Pi Zero as the central hub. (d) The system uses a standard Wi-Fi router to facilitate communication between the sensor nodes and the hub, enabling real-time data transmission and visualization. (e) The web-based interface allows users to visualize and record motion data in real-time.}
  
  \label{fig_system_overview}
\end{figure*}

Inspite of these advancements, position estimation remains a significantly harder problem than orientation. It typically requires double integration of accelerometer signals, which amplifies noise and leads to rapid drift over time \cite{foxlin2005, sabatini2011}. In the absence of external spatial references (e.g., GPS, beacons, or vision-based tracking), pure inertial positioning becomes unreliable over even short durations.

To overcome this, the Mesquite system adopts a hybrid approach by anchoring the user’s hip position using WebXR World Mapping \cite{webxr, webxr-wm}, which leverages computer vision SLAM techniques (e.g., Visual Odometry, sensor fusion, on-board ML modeling) from a hip-mounted smartphone to establish an absolute spatial frame. The rest of the body bones are tracked using 15 body-worn IMUs for orientation, and joint positions are estimated through forward kinematics, reducing the need for full-body positional tracking. This design choice exploits the fact that for many applications, accurate tracking of the hip position is sufficient, as the rest of the skeleton can be reconstructed using only bone orientations \cite{gonzalez2020}. In order to capture a full motion frame we only need the position/orientation of the hip and the orientation of the rest of the bones of the body. This approach allows for a more practical and cost-effective solution while maintaining high accuracy in motion capture. This data is exported in BVH format, which is widely used in animation and motion analysis software. The raw data of each sensor is also logged for further analysis and post-processing if needed.

\section{System Design}

The complete Mesquite system includes (see Fig. \ref{fig_system_overview}):

\begin{enumerate}
  \item \textbf{15 WiFi-capable IMU pods:} Attached to body segments for orientation tracking.
  \item \textbf{Android smartphone:} Positioned at the hips for camera-based spatial anchoring.
  \item \textbf{Raspberry Pi Zero:} Acts as a USB dongle and local HTTP and WebSocket server.
  \item \textbf{Standard Wi-Fi router:} Facilitates data transport between devices.
  \item \textbf{Web-based interface:} Accessible\footnote{at \url{https://mesquite.cc} and source available at \url{https://github.com/Mesquite-Mocap/mesquite.cc}} using any Web Serial-compatible browser for visualization and recording.
\end{enumerate}

All aspects of the system are open-source, designed with modularity and affordability in mind. By combining embedded sensor fusion with modern web technologies such as WebSockets, WebSerial and Progressive Web Apps, Mesquite demonstrates a practical and extensible solution to the longstanding challenges of full-body motion capture. The system is designed to be modular and extensible, allowing users to customize the number of sensors and their placement on the body. Custom sensor configurations can be easily implemented to accommodate different use cases, user preferences and skeletons. The system consists of three main components: hardware, software, and data processing (Fig. \ref{fig_system_overview}), which are described in detail below.

\subsection{Hardware Components}

\begin{figure}[h]
  \centering
  \includegraphics[width=3.2in]{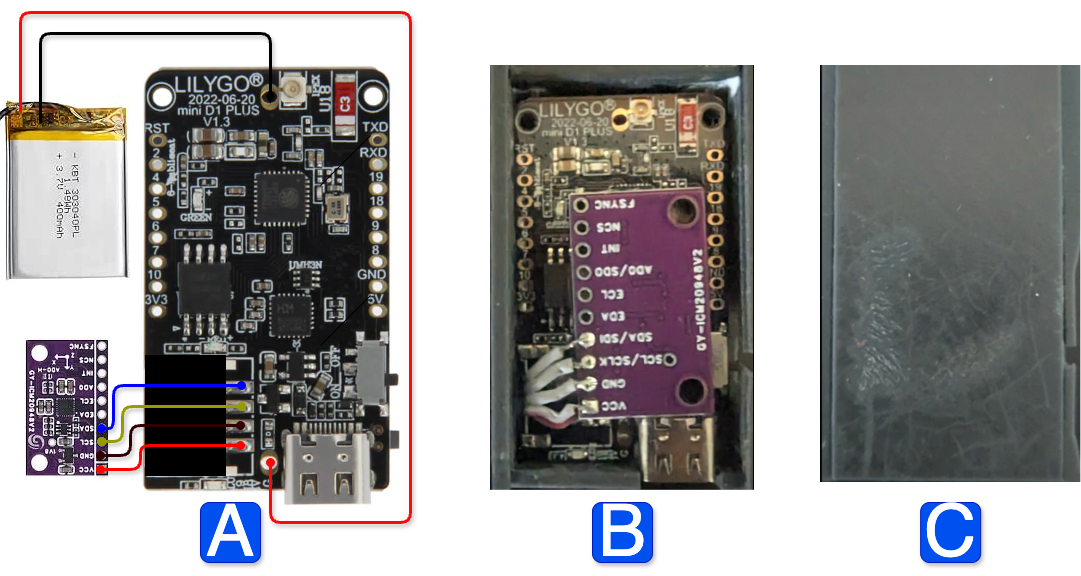}
  \caption{The Mesquite sensor node (``pod'') (a) i2c ICM20948 + ESP32 C3 (TTGO T-OI) + LiPo battery (b) pod in 3d-printed enclosure (c) Full pod in enclosure}
  \label{fig_hardware}
  \end{figure}

\begin{figure}[h]
  \centering
  \includegraphics[height=4in]{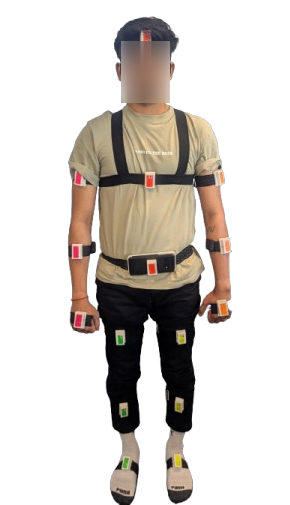}
  \caption{All 15 pods are attached to the body using adjustable straps. The system is designed to be lightweight, unobtrusive, and one size fits all. The smartphone is mounted on the hip for spatial anchoring.}
  \label{fig_tpose}
  \end{figure}

The system consists of 15 wearable devices (``pods") that attach to the user's body via adjustable elastic straps (Fig. \ref{fig_tpose}). Each node contains a 9-axis ICM20948 IMU sensor combining accelerometer, gyroscope, and magnetometer; a dual-core ESP32 microcontroller operating at 240MHz; a 400mAh lithium-polymer battery (Fig. \ref{fig_hardware} (a)); and a 3D-printed enclosure designed for comfort and durability ((Fig. \ref{fig_hardware}) (b,c)). We do not use the magnetometer for regular operation, but it can be logged for post-processing if needed. The ESP32 microcontroller is responsible for data acquisition, preprocessing, and wireless communication. The ICM20948 IMU samples at 1000Hz, while the ESP32 processes the data at 100Hz using its integrated Digital Motion Processor (DMP) to generate quaternion orientation data. The system is powered by a rechargeable lithium-polymer battery, providing approximately 4-5 hours of continuous operation (at 30 fps).

  The sensor nodes are strategically placed on key body segments to capture a complete skeletal representation: head, upper arms (×2), forearms (×2), hands (×2), hips, spine, thighs (×2), calves (×2), and feet (×2). The system design is modular, allowing users to deploy fewer nodes when only specific body segments require tracking. The nodes are attached using adjustable straps (Fig. \ref{fig_tpose}), ensuring a secure fit during movement. The pods can attach to the strap via an elastic loop, velcro or magnets. The system is designed to be lightweight and unobtrusive, allowing users to perform a wide range of activities without hindrance.

The ESP32 microcontroller implements a dual-task FreeRTOS architecture, with one core dedicated to sensor data acquisition and preprocessing, while the second core handles wireless communication. This separation ensures consistent sampling rates regardless of network conditions.
The central hub consists of a Raspberry Pi Zero W functioning as a network coordinator and data aggregator. This component receives data from all sensor nodes, performs initial processing, and serves the web interface to the hipworn phone.

\begin{figure}[h]
  \centering
  \includegraphics[height=2in]{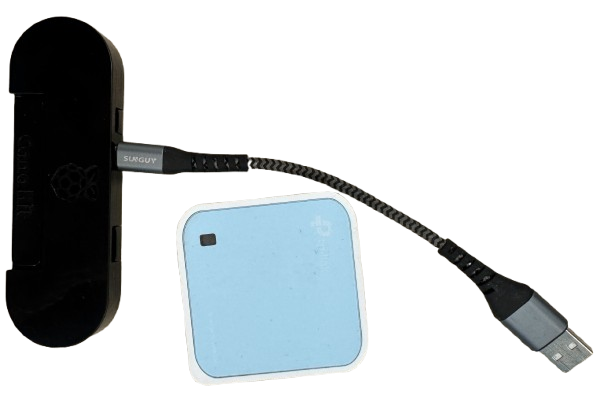}
  \caption{Mesquite dongle (Raspberry Pi Zero W Wi-Fi module) and Wifi router. The Raspberry Pi Zero W acts as a USB dongle and local HTTP and WebSocket server, while the Wi-Fi router facilitates communication between the sensor nodes and the hub.}
  \label{fig_dongle}
  \end{figure}

\begin{figure}[h]
    \centering
    \includegraphics[height=2.8in]{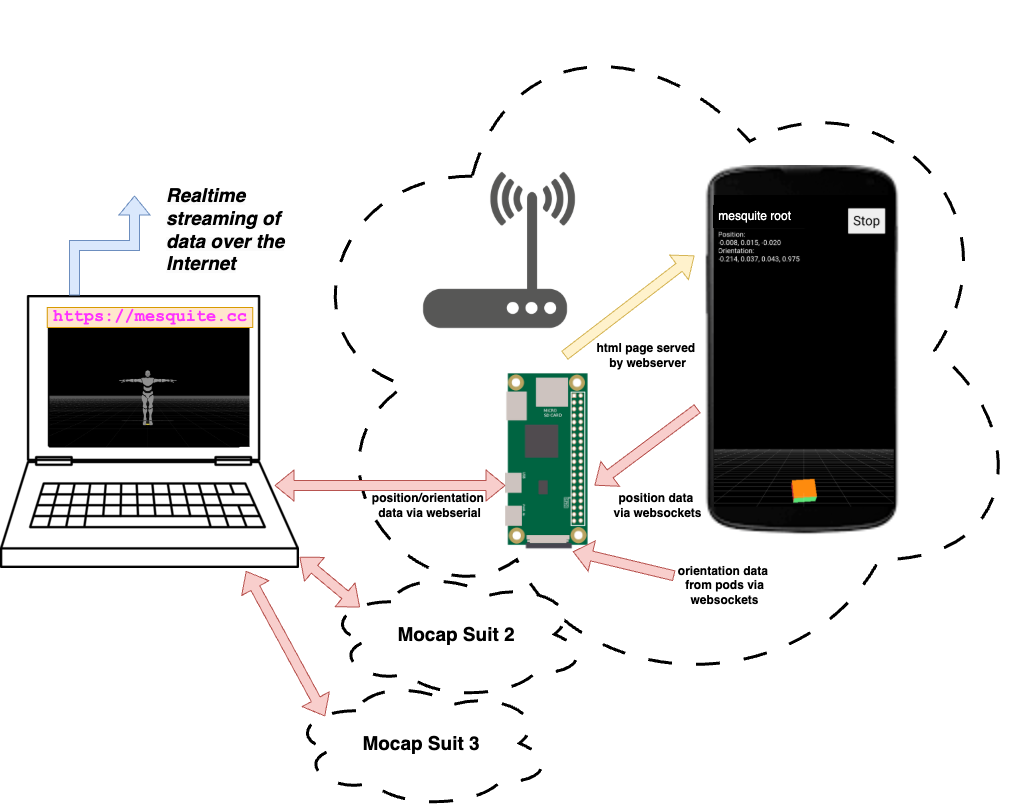}
    \caption{WebXR SLAM: The smartphone uses WebXR World Mapping to establish an absolute spatial frame, which is then used to track the user's hip position. The system uses a standard Wi-Fi router to facilitate communication between the sensor nodes and the hub, enabling real-time data transmission and visualization.}
    \label{fig_webxr-slam}
    \end{figure}

  The Raspberry Pi Zero W works as a dongle, while connected to the local Wi-Fi network and acting as a USB device for data aggregation. It runs a Node.js application that serves the web page for the phone and handles the positioning data received from the phone and the orientation data from the pods (via websockets; see Fig. \ref{fig_webxr-slam}). The Raspberry Pi Zero W is powered by a 5V power supply (when connected as a USB device) and uses a built-in Wi-Fi module to connect to the local network. The system can also be configured to work as a standalone access point, allowing it to operate without external network dependencies (if less than 15 pods are used). Adding a dedicated router to the MoCap system enables the use of multiple mocap suits in the same space without interference. Also the computer running the webapp can be connected to any Wi-Fi network, thereby having an independent internet connection. This allows for streaming of mocap data in real time to any other computer on the Internet, or to a cloud server for further processing and integration with other applications.

\subsection{Software Components}

Unlike simpler peer-to-peer approaches, Mesquite employs a robust dongle which serves both the webxr SLAM HTML (to the hip phone) and the websocket server that aggregates the data, with a standard Wi-Fi router serving as the communication infrastructure (See Figures~\ref{fig_system_overview} and \ref{fig_webxr-slam}). This configuration provides greater range and reliability compared to Bluetooth or direct ESP32-to-ESP32 communication; the ability to handle multiple sensors simultaneously without data congestion; simplified security implementation through standard Wi-Fi encryption; and compatibility with infrastructure widely available in most environments. This configuration also works offline and in-the-field as it does not depend on any external network infrastructure/Internet. The system can be set up in a matter of minutes, with no need for complex configuration or installation processes. The webapp is accessible via any modern web browser and can be installed as an offline app. The sensor nodes connect to the local Wi-Fi network using the ESP32's integrated wireless capabilities, with the Raspberry Pi serving as the central data sink. The dongle is assigned a static IP which makes the system easy to set up and use. 

\begin{figure}[h]
  \centering
  \includegraphics[width=3.2in]{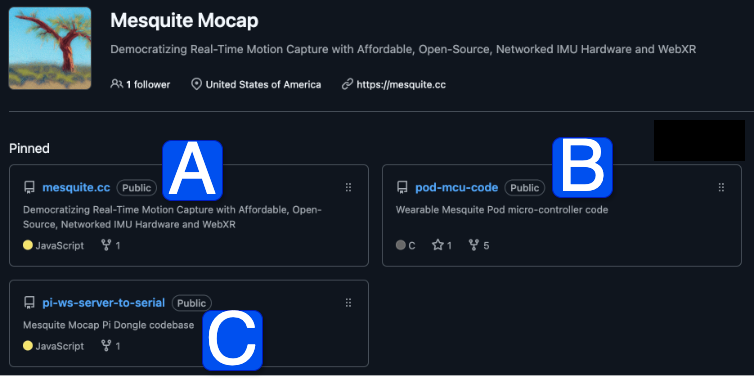}
  \caption{Three main software repositories for the Mesquite system: (a) Web interface for visualization and recording, (b) Sensor node firmware for data acquisition and preprocessing, and (c) Central hub software for data aggregation and synchronization.}
  \label{fig_software}
  \end{figure}

The software architecture consists of three primary components: 
\begin{enumerate}[label=(\alph*)]
  \item \textbf{The Webapp:} A web-based application that provides real-time visualization and recording of motion data\footnote{\url{https://github.com/Mesquite-Mocap/mesquite.cc}}. The web interface is built using modern web technologies such WebSerial, PWA and WebGL, allowing for real-time rendering and offline use. It is designed to be responsive and compatible with various devices, including smartphones, tablets, and desktop computers (Fig. \ref{fig_software} (a)). The web interface allows users to visualize the motion data in real-time, record the data for later analysis, and export the data in bvh and csv formats. The web interface is built using Three.js, providing a smooth and interactive user experience. The web interface also includes a calibration tool workflow (discussed next) that allows users to calibrate the system before use.
  \item \textbf{Sensor Node Firmware:} The firmware running on the ESP32 microcontroller, responsible for data acquisition, preprocessing, and wireless communication\footnote{\url{https://github.com/Mesquite-Mocap/pod-mcu-code}}. The code is written in C++ and uses the Arduino framework for ease of development. The firmware implements a dual-task FreeRTOS architecture, with one core dedicated to sensor data acquisition and preprocessing, while the second core handles wireless communication. This separation ensures consistent sampling rates regardless of network conditions (Fig. \ref{fig_software} (b)). This repository also contains 3D STL files for all enclosures and the calibration plate.
  \item \textbf{Central Hub Software:} The software running on the Raspberry Pi Zero W, responsible for data aggregation, synchronization, and serving the web interface\footnote{\url{https://github.com/Mesquite-Mocap/pi-ws-server-to-serial}}. The software is run on the Raspberry Pi Zero W, which acts as a USB dongle and local HTTP and WebSocket server. The README provides detailed instructions for setting up the Raspberry Pi as a dongle, including installing the required dependencies and configuring the network settings. The software is written in Node.js and uses the Express framework for serving the web interface and WebSocket for real-time data transmission (Fig. \ref{fig_software} (c)). The README incudes how to automate the running of this software on boot, so that the Raspberry Pi can be used as a standalone plug-and-play dongle.

\end{enumerate}

The firmware implements a full sensor fusion algorithm on the ESP32 to generate quaternion orientation data, leveraging the ICM20948's Digital Motion Processor (DMP) for offloading computationally intensive operations. This edge computing approach significantly reduces wireless bandwidth requirements by transmitting only quaternion data rather than raw sensor values.
The hub software manages connections with all sensor nodes, handles data synchronization, and implements the hierarchical skeletal model used for visualization. WebSocket technology enables real-time data streaming to the browser interface with minimal latency.

\subsection{Calibration Process}

\begin{figure}[h]
  \centering
  \includegraphics[width=3.2in]{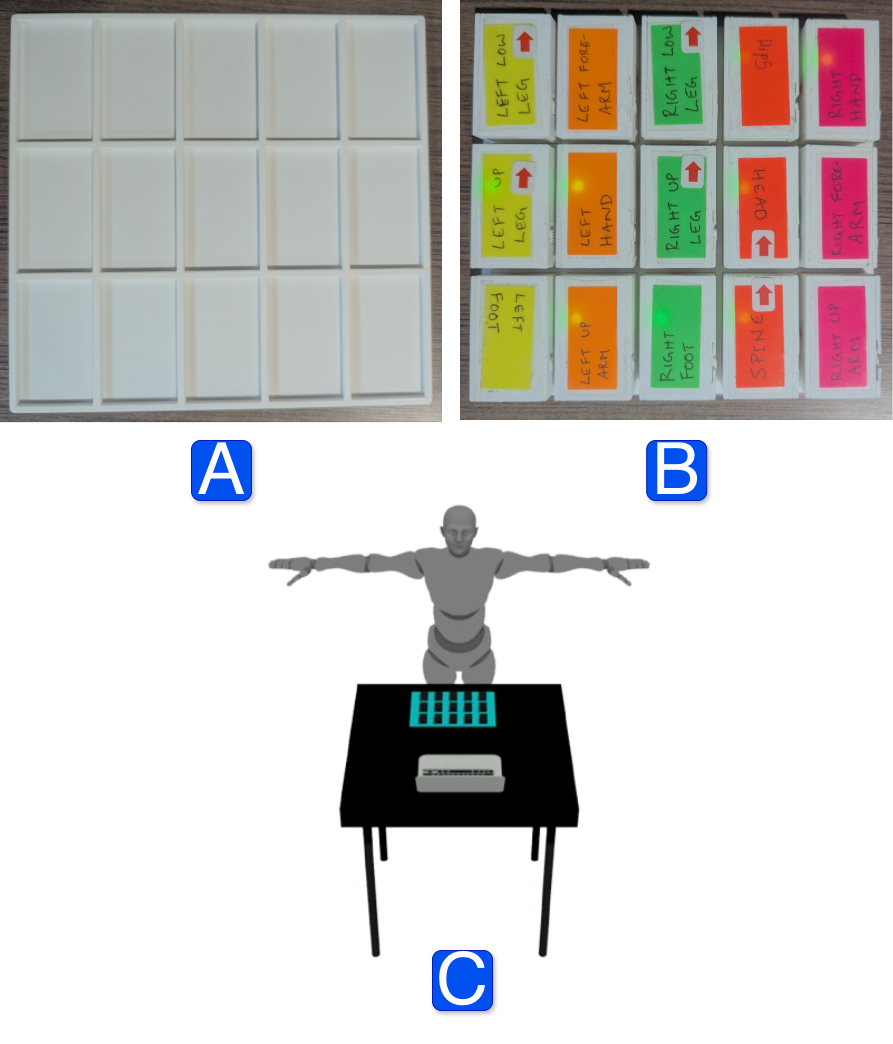}
  \caption{Calibration plate: (a) Empty calibration plate. (b) Calibration plate with pods (c) Calibration plate placement w.r.t t pose}
  \label{fig_calibration}
  \end{figure}

Mesquite implements a straightforward calibration protocol to ensure accurate motion capture. Users first place the sensor nodes on a flat surface, such as a table, to establish a reference frame. The calibration plate (Fig. \ref{fig_calibration}) guides users through the process of aligning the sensor nodes with the reference frame. The UI allows the users to do a ``deep'' calibration, where the user can place the pods on a flat surface and press the ``box calibrate'' button. The system then uses the accelerometer data to determine the orientation of each pod relative to the reference frame. This process is essential for ensuring that the sensor nodes are correctly aligned with the user's body segments during motion capture. These box calibration values can be downloaded as a json and uploaded to the webapp. 

After the box calibration is done or previous values are uploaded, the user can wears the pods and presses the ``T-Pose'' button. These two values ($q_{box}$ and $q_{tpose}$) are used to calculate the transformation matrix for each pod. The system then uses this transformation matrix to convert the sensor data from the local coordinate system of each pod to a global coordinate system. This process is essential for ensuring that the sensor data is accurately aligned with the user's body segments during motion capture.

Each bone then uses the raw data ($q_{raw}$) and the ${q_{tpose}}$ and $q_{box}$ values to calculate the final quaternion data ($q_{local}$) for each pod. The transformation matrix is calculated as follows:

\begin{equation}
q_{local} = q_{raw} \cdot {q_{tpose}}^{-1} \cdot q_{box}
\end{equation}

where $q_{raw}$ is the raw quaternion data from the sensor node, $q_{tpose}$ is the transformation matrix for the T-Pose calibration, and $q_{box}$ is the transformation matrix for the box calibration. The system then uses this transformation matrix to convert the sensor data from the local coordinate system of each pod to a global coordinate system. This process is essential for ensuring that the sensor data is accurately aligned with the user's body segments during motion capture.

The final mapping to the skeleton involves accomadating the parent bone quaternions and the local bone quaternions. The final quaternion data for each pod is calculated as follows:
\begin{equation}
q_{final} = {q_{parent}}^{-1} \cdot q_{local}
\end{equation}

where $q_{parent}$ is the quaternion data for the parent bone, and $q_{local}$ is the quaternion data for the local bone. 

  This calibration process typically requires less than two minutes and automatically compensates for minor variations in sensor placement between sessions. The procedure and the UI is designed for one person to perform without assistance, making it suitable for both individual and group settings. The calibration process is designed to be user-friendly and intuitive, with clear instructions and visual feedback provided throughout the process. The system also includes a ``Start Over" button that allows users to recalibrate the system at any time during the session. This feature is particularly useful in dynamic environments where sensor placement may change or when users need to switch between different calibration setups. Also T-pose calibration can be performed in between captures (which only takes 5 seconds) to compensate for any drift that may occur during the session.

\begin{figure}[h]
  \centering
  \includegraphics[height=2.5in]{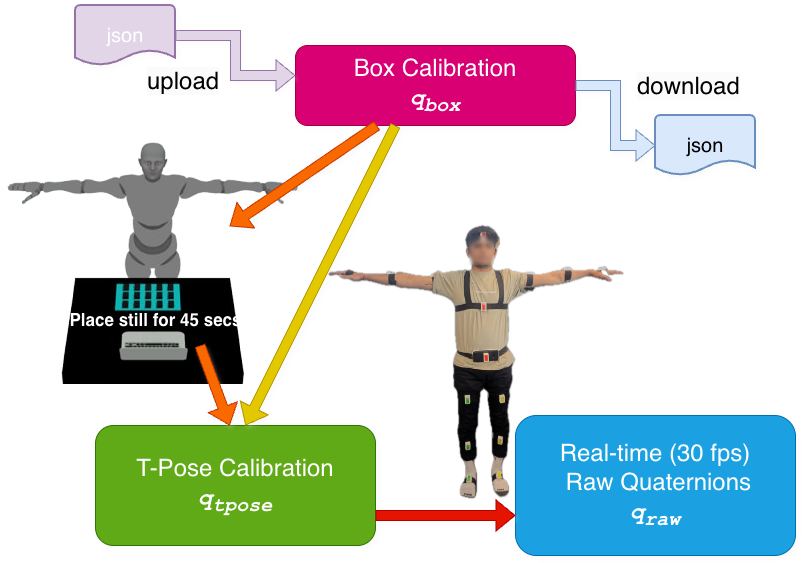}
  \caption{Calibration process}
  \label{fig_calibration-flow}
  \end{figure}

\subsection{Skeletal Model and Animation}

Sensor data undergoes multiple transformations to create accurate skeletal animations. The ICM20948 samples accelerometer, gyroscope, and magnetometer at 1000Hz. Edge processing occurs as the ESP32 accesses the ICM20948's DMP to generate quaternion orientation data at 100Hz. After downsampling, data is packaged  WebSocket packets at 32Hz for transmission wirelessly to the central hub. Reference frame conversion transforms data from sensor-local to global coordinate systems, enabling hierarchical skeleton reconstruction through applying quaternion rotations to the skeletal model. Finally, Kalman filtering and SLERP are applied for noise reduction and smooth animations.

The skeletal model is represented as a hierarchical structure, with each joint defined by a parent-child relationship. The system uses a simplified kinematic model with 15 joints corresponding to the sensor node placements. Each joint's orientation is derived from the quaternion data transmitted by the respective pod (corresponding to a bone), and the global orientation of each joint is calculated based on its parent's joint position and orientation.

The skeletal model is visualized in real-time using WebGL, with the browser interface rendering the 3D animation based on the processed data. To improve tracking accuracy and mitigate common IMU limitations, the system employs multiple filtering approaches. Kalman filtering is implemented on the ESP32 to reduce noise in orientation estimation and mitigate gyroscope drift \cite{welch2006}.  Spherical linear interpolation (SLERP) is applied for smooth transitions between quaternion states \cite{shoemake1985}. The system does not require magnetometer data for regular operation, making it robust to magnetic interference commonly found in indoor environments. However, magnetometer readings can optionally be incorporated to prevent long-term heading drift in extended recording sessions.

\subsection{ Data Export and Interoperability}
The system exports motion data in the Biovision Hierarchy (BVH) format, enabling interoperability with existing animation and analysis software. The export pipeline includes quaternion to Euler angle conversion for compatibility with standard BVH implementations, joint hierarchy definition matching common animation standards, frame data export at configurable frame rates (15-45 FPS), and metadata inclusion for raw sensor data. The exported BVH files are compatible with software including Blender\cite{blender}, Maya\cite{maya}, Unity\cite{unity}, Unreal Engine\cite{unreal}, and other specialized biomechanical analysis tools. The system also supports CSV export for raw data analysis, allowing users to access the underlying sensor data for custom processing or visualization. This flexibility enables researchers and developers to integrate Mesquite into their existing workflows and leverage the system's capabilities for a wide range of applications.

Real-time streaming of JSON data is also supported, allowing users to access the raw sensor data for custom processing or visualization. This flexibility enables researchers and developers to integrate Mesquite into their existing workflows and leverage the system's capabilities for a wide range of applications.

\subsection{Other Considerations}

Each sensor node incorporates intelligent power management features to maximize battery life and operational duration. Deep sleep modes activate during periods of inactivity. The straps are made of elastic material to ensure a secure and comfortable fit during movement. All 3d printed parts are designed for durability and ease of assembly .These features enable approximately 4-5 hours of continuous comfortable operation before requiring recharging. The system includes LED indicators for battery status and USB-C charging ports for convenient recharging.

\section{Evaluation}

To validate the accuracy of Mesquite, we conducted comparative benchmarking tests against an OptiTrack optical motion capture system\cite{optitrack}, widely considered the gold standard in motion tracking technology. The benchmarking methodology included simultaneous capture of a variety of human movements ranging from simple actions (walking, reaching) to complex motions (climbing, dancing etc).

\subsection{Methodology}
Our validation procedure followed these steps:
\begin{enumerate}
    \item A participant was equipped with both Mesquite sensors and OptiTrack markers.
    \item A series of standardized movements was performed, including:
    \begin{itemize}
        \item Walking in a straight line,
        \item Walking in circles,
        \item Climbing stairs,
        \item 90-degree turns,
        \item Arm reaching patterns,
        \item Squatting motions, and
        \item Dynamic dance movements.
    \end{itemize}
    \item Data was captured simultaneously from both systems.
    \item Joint angles were calculated and compared.
    \item Position tracking was assessed for drift and accuracy.
\end{enumerate}

The benchmarking process involved simultaneous data capture from both Mesquite and OptiTrack systems. Joint angles were calculated and compared to assess angular accuracy, while position tracking was evaluated for drift and consistency.

\subsection{Results}

\begin{figure*}[p]
  \centering
  \includegraphics[width=.9\textwidth]{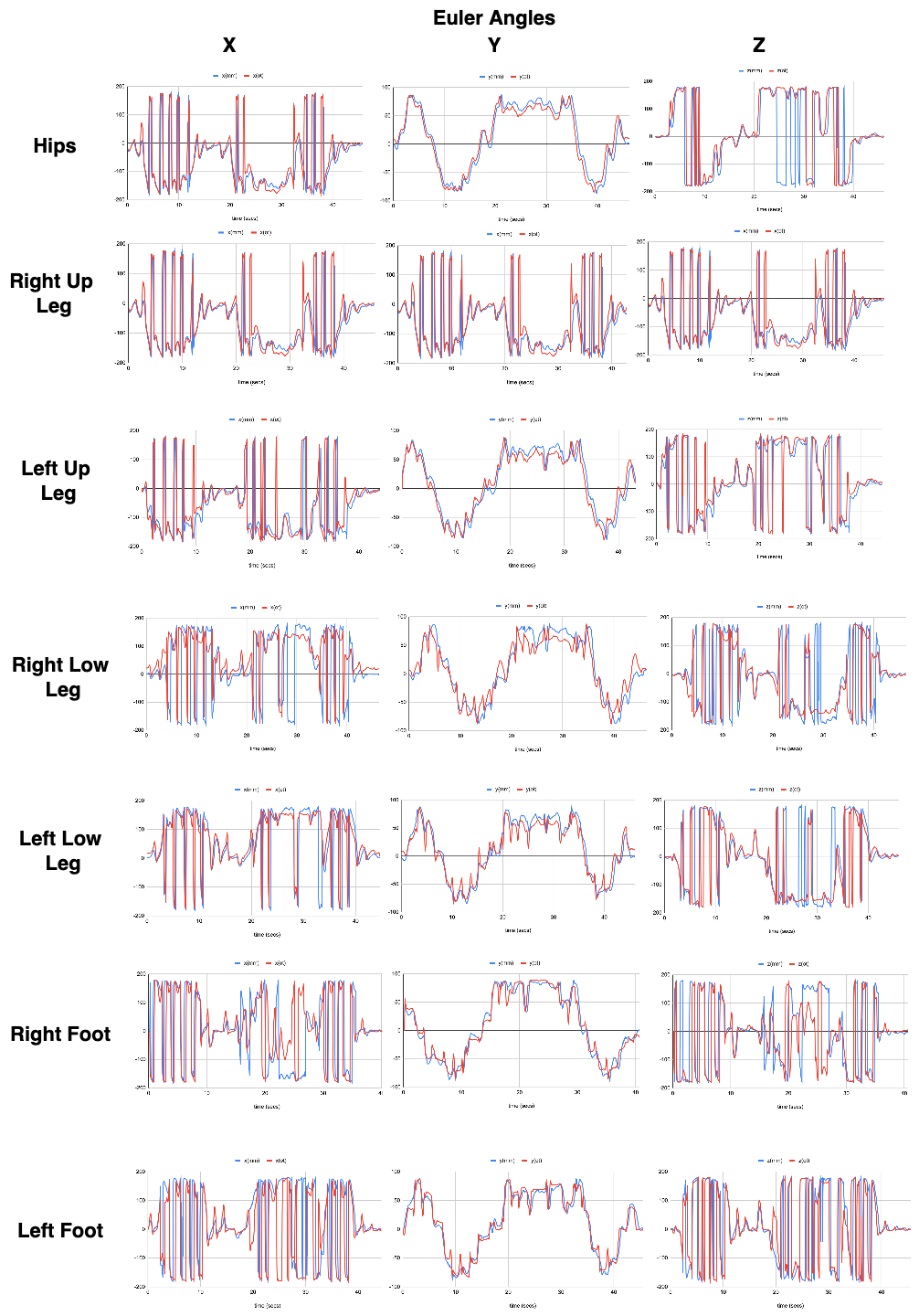}
  \caption{Benchmarking results comparing Mesquite Mocap with OptiTrack system.}
  \label{fig_results1}
\end{figure*}


\begin{figure*}[p]
  \centering
  \includegraphics[width=.8\textwidth]{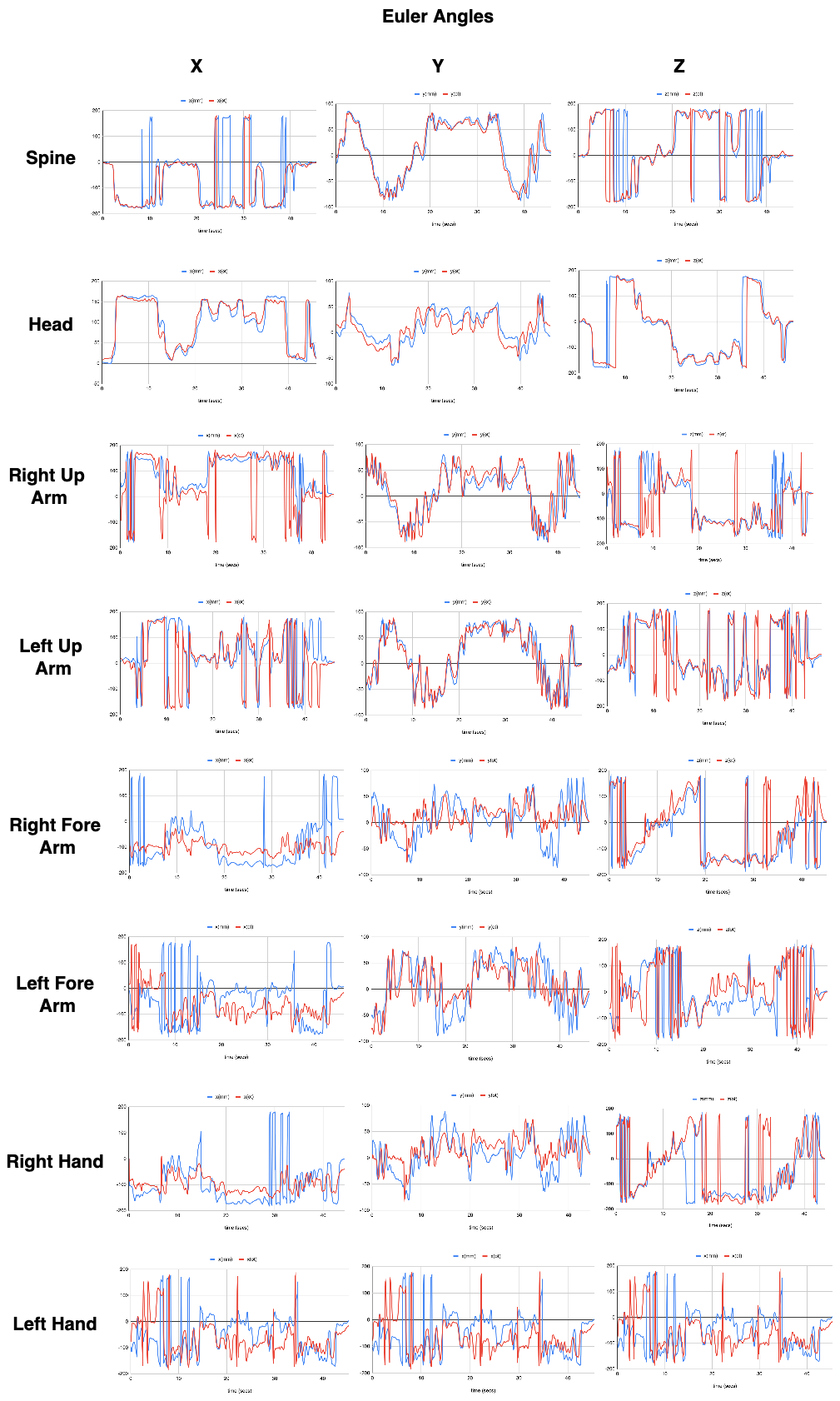}
  \caption{Benchmarking results comparing Mesquite Mocap with OptiTrack system (contd.).}
  \label{fig_results2}
\end{figure*}

The results demonstrate that Mesquite achieves angular accuracy within 2-5 degrees of the OptiTrack system for most joint rotations, with particularly strong performance in larger body segments (Figures~\ref{fig_results1} and \ref{fig_results2}). Position tracking showed minimal drift over 10-minute sessions, with average positional errors remaining below 5cm in dynamic movements. These results confirm that Mesquite delivers professional-grade motion capture capabilities at a fraction of the cost, making it an accessible solution for researchers, educators, and developers.

Detailed benchmarking data, comparative animations, and BVH file examples from both systems are available at \url{https://mesquite.cc/benchmarks}. These resources provide visual and quantitative evidence of Mesquite's performance relative to systems costing 20 times more, validating its suitability for professional applications.

These results confirm that Mesquite delivers professional-grade motion capture capabilities at a fraction of the cost, making it an accessible solution for researchers, educators, and developers (See Figures~\ref{fig_results1} and \ref{fig_results2}).

\section{Applications}

While not designed for a single use case, Mesquite can add value across multiple professional domains:

\begin{itemize}
  \item \textbf{Biomechanics Research:} Researchers can leverage Mesquite's affordability and portability in their work. The system enables gait analysis in both clinical and field settings, while also facilitating detailed joint kinematics studies during complex movements. Longitudinal movement research has been enhanced through home-based participant monitoring, and workplace ergonomic assessments benefit from the system's non-intrusive deployment capabilities.
  \item \textbf{Animation and Game Development:} Creative professionals can use Mesquite to enhance their projects through efficient character animation for indie games and films. The technology also enables live performance capture for interactive experiences, bridging the gap between performers and digital environments.
  \item \textbf{Clinical Applications:} Healthcare professionals have found significant value in Mesquite's capabilities. Physical therapists monitor patient progress with objective movement data, while rehabilitation programs track exercise adherence more effectively. The system provides quantitative assessment of movement disorders and enables remote patient monitoring, expanding access to quality care beyond traditional clinical settings.
  \item \textbf{Sports Performance Analysis:} Coaches and athletes utilize Mesquite to gain competitive advantages through comprehensive technique analysis and optimization. Training progress can be quantified with precise metrics, while potential injuries are prevented through early identification of problematic movement patterns. Comparative analyses between athletes provide valuable insights for team development and individual improvement strategies.
  \item \textbf{Education:} Educational institutions integrate Mesquite across various disciplines to enhance learning experiences. Anatomy and biomechanics courses utilize the technology for practical demonstrations, while animation and game design programs provide students with professional-grade tools. Human-computer interaction research benefits from the system's spatial tracking capabilities, and computer science projects involving spatial computing gain a practical implementation platform. The system has already been adopted by several universities for teaching and research purposes.
\end{itemize}

\section {Future Work and Development}

Future enhancements planned for Mesquite include:
\begin{itemize}
  \item \textbf{Improved Frame Rate:} Hardware and firmware optimizations to achieve 60+ FPS for capturing faster movements.
  \item \textbf{Hand Tracking:} Additional specialized sensors for fingers expressions.  
  \item \textbf{Face Tracking:} Additional cameras for and facial expression tracking.
  \item \textbf{Sensor Reduction:} Development of a diffusion model to reduce the required number of sensors from 15 to 6 while maintaining accuracy through machine learning interpolation.
  \item \textbf{Neural Network Integration:} Implementation of a Recurrent Neural Network (RNN) for gesture recognition, movement classification, and predictive filtering.
  \item \textbf{Cloud Platform:} Secure cloud storage and analysis tools for collaborative research and remote monitoring applications.
  \item \textbf{Mixed Reality Integration:} Direct streaming to VR/AR platforms for immersive applications.
\end{itemize}

These developments will be pursued through both core team efforts and community contributions, maintaining the open-source nature of the project.

\section{Conclusion}
Mesquite represents a significant step toward democratizing access to motion capture technology. By combining affordable hardware with open-source software and real-time visualization capabilities, the system enables applications previously constrained by cost and technical barriers. The flexible and extensible nature of Mesquite makes it a valuable tool for researchers, educators, healthcare professionals, and content creators seeking accurate motion data without prohibitive investment.
Our benchmarking confirms that the system achieves accuracy levels suitable for most professional applications at a fraction of the cost of traditional solutions. The PWA webapp eliminates software installation barriers, while the wireless nature of the system removes physical constraints of tethered alternatives. By making Mesquite fully open-source, we aim to foster a community of users and developers who can continue to improve the system and adapt it to new applications. This collaborative approach ensures that motion capture technology can evolve to meet the needs of a diverse user base beyond the traditional entertainment industry focus. The combination of affordability, accuracy, and accessibility positions Mesquite as a transformative tool in fields ranging from healthcare to education to independent media creation. As motion analysis becomes increasingly valuable across disciplines, Mesquite provides a democratized pathway to implementing this technology without the traditional financial barriers.

\bibliographystyle{IEEEtran}
\bibliography{main}



\newpage

\end{document}